\documentclass{appolb}
\usepackage[dvips]{graphicx} 
\usepackage{epsfig,rotate} 
\usepackage{color}
\usepackage{psfrag}
\usepackage{amsmath,amstext,amsbsy,amsopn,amsfonts,amssymb} 

\newcommand{\ft}[2]{{\textstyle\frac{#1}{#2}}}
\newcommand\lr[1]{{\left({#1}\right)}}

\begin{document} 
\bibliographystyle{unsrt}
   
\title{
 Cusp anomalous dimension   in maximally supersymmetric  Yang-Mills theory 
\thanks{
Seminar talk presented at the XLVIII Cracow School of Theoretical Physics: ``Aspects of Duality 2008'', Zakopane, Poland, June 13-22, 2008.}
\author{{\bf Jan Kotanski}
\address{II.~ Institute Theoretical Physics, Hamburg University, Germany\\
M. Smoluchowski Institute of Physics, Jagellonian University,
Poland}}}
                                 
\date{October 14, 2008} 
\maketitle 
 
\begin{abstract} 
The main features of the cusp anomalous dimension  in ${\cal N}=4$ supersymmetric  Yang-Mills theory are reviewed. Moreover, the strong coupling expansion of the cusp derived in Ref.~\cite{Basso:2007wd} is presented.
\end{abstract} 


\section{The cusp and  AdS/CFT correspondence}

This talk is mainly based on the work \cite{Basso:2007wd} performed in collaboration with B. Basso\footnote{Laboratoire de Physique Th\'eorique, Universit\'e de Paris XI,
France} and  G. Korchemsky\footnotemark[1], where the method of  strong coupling 
expansion of the  cusp anomalous dimension was found.
The cusp anomalous dimension \cite{Polyakov:1980caKorchemsky:1985xjKorchemsky:1988si}, called also the cusp, is a physical observable, which appears 
in many branches
of particle physics. It is related with the logarithmic 
growth of the anomalous
dimensions of high-spin Wilson operators
and with the gluon Regge trajectory,
it governs  behavior of the Sudakov form factors as well as
it defines  infrared singularities of on-shell scattering amplitude.

In order to define the cusp anomalous dimension in
in ${\cal N}=4$ supersymmetric Yang-Mills (MSYM) theory
one can consider local operators composed of $L$ scalar fields
and $S$ covariant derivatives, 
$\langle D^{s_1}X D^{s_2}X \ldots D^{s_L}X \rangle$, with the spin
$S=\sum_{k=1}^L s_k$.
In the high spin limit the anomalous dimensions of these operators read as
\begin{equation}
\gamma_{S}^{(L=2)}(g) =2 \Gamma_{\rm cusp}(g) \ln S+ \ldots, 
\end{equation}
where the leading coefficient, $\Gamma_{\rm cusp}(g)$ is called cusp
anomalous dimension. 
In ${\cal N}=4$ SYM theory the cusp is a fundamental quantity
and it depends neither on the twist L\footnote{For higher twist there are additional degrees of freedom and the cusp defines the high spin asymptotic of the minimal anomalous dimension}
nor on composed fields.

%
In 1998 the AdS/CFT correspondence was proposed by  
Maldacena, Polyakov, Klebanov, Gubser and  Witten
\cite{Maldacena:1997reGubser:1998bcWitten:1998qj}.
It relates MSYM operators to 
th IIB string theory observable, 
\ie the cusp anomalous dimension in MSYM theory corresponds to   
energy of folded strings rotating  in AdS$_3$
\cite{Gubser:2002tv}.
Due to  dual properties of the AdS/CFT conjecture
calculations of the strong coupling  limit in the SYM theory
are equivalent to the semiclassical expansion of the string theory.
Moreover,  if
the physical quantity like the cusp can be
calculated in ${\cal N}=4$ SYM theory
both in the weak as well as strong coupling expansion, \eg making use of
integrable methods \cite{Belitsky:2004czBeisert:2004ry}, 
an additional aspect of investigation arises.
In this special case 
one may try to test a validity of the AdS/CFT correspondence.

From the string theory side the cusp anomalous dimension
\begin{equation}
\Gamma_{\rm cusp}(g) = 2 g - \frac{3\ln 2}{2\pi} + O(1/g)\,,\qquad g =
\frac{\sqrt{\lambda}}{4\pi},
\end{equation}
with $\lambda = g^2_{\rm YM} N_c$ being 't Hooft coupling
was calculated in 
the 1-loop string perturbation calculation
\cite{Gubser:2002tv,Frolov:2002avKruczenski:2002fb}
and from the string Bethe ansatz 
\cite{Casteill:2007ct}.
The question is if this result can be also obtained from
${\cal N}=4$ SYM theory.

\section{Beisert-Eden-Staudacher equation }

In Ref.~\cite{Eden:2006rxBeisert:2006ez}  Beisert, Eden  and Staudacher derived from all-loop Bethe ansatz \cite{Arutyunov:2004vxStaudacher:2004tkBeisert:2005fwJanik:2006dcBeisert:2005cwHernandez:2006tkBeisert:2006ib} the equation
\begin{equation}
\widehat\sigma_g (t) = \frac{t}{e^{t}-1}\left[K_g(2gt, 0) 
- 4g^{2}\int_{0}^{\infty}
dt' K_g(2gt, 2gt') {\widehat\sigma_g(t')}\right],
\end{equation}
with a complicated kernel, 
$K_g(t,t')=\sum_{n,m=
1}^\infty z_{nm}(g) \frac{J_n(t) J_m(t')}{tt'}$,
defined in Ref.~\cite{Eden:2006rxBeisert:2006ez}.
The fluctuation density $\widehat\sigma_g (t)$ is related 
to the Fourier transform
of the  Bethe roots distribution 
\cite{Korchemsky:1995beBelitsky:2006en}
and it predicts the cusp anomalous dimension 
for arbitrary values of the coupling
constant $\Gamma_{\rm cusp}(g)=8 g^2  \widehat \sigma_g (0)$.
The weak coupling expansion obtained from the BES equation is given as
\begin{eqnarray}
2 \Gamma_{\rm cusp}(g)&=&8 {g^2} - \ft{8}{3}\pi^2 {g^4}
+\ft{88}{45} \pi^4 {g^6}
-16\left(\ft{73}{630} \pi^6 +4 \zeta(3)^2 \right){g^8}+ \nonumber \\
&& + 32\left(\ft{887}{14175} \pi^8+\ft{4}{3} \pi^2 \zeta(3)^2
+40 \zeta(3) \zeta(5) \right){g^{10}}
+
\nonumber \\
&&  -64\left( \ft{136883}{3742200} \pi^{10}+
\ft{8}{15} \pi^4\ \zeta(3)^2+ \ft{40}{3} \pi^2 \zeta(3) \zeta(5)
\right. \nonumber \\&& \left.
+210 \zeta(3) \zeta(7) 
+102 \zeta (5)^2\right){g^{12}} +\ldots,
\end{eqnarray}
a sign alternating series with the convergence radius $\ft{1}{4}$.
The three loop result was also obtained
from QCD using the maximal tanscendentality principle
\cite{Kotikov:2006tsVogt:2004mw}
and the four loop value was checked in perturbation calculations
\cite{Bern:2006ewCachazo:2006az}.

\section{Strong coupling expansion in numerical approach}

\begin{figure}
\begin{center}
\begin{picture}(280,120)
\put(40,0){\epsfysize4.5cm \epsfbox{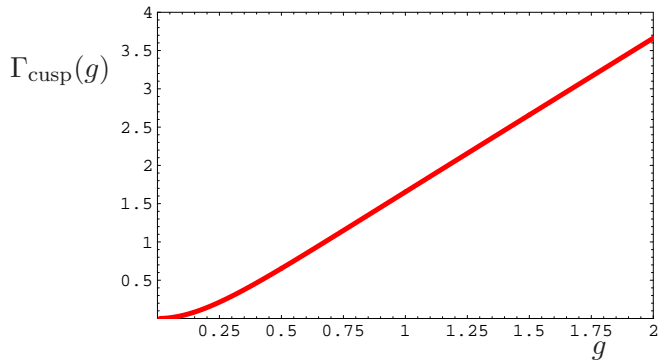}}
\put(0,100){$\Gamma_{\rm cusp}(g)$}
\put(220,-5){$g$}
\end{picture}
\end{center}
\caption{The cusp anomalous dimension as a function of coupling constant g.}
\label{fig:cusp}
\end{figure}

In order to solve the BES equation numerically 
\cite{Benna:2006nd}
one can 
expand the fluctuation density over Bessel functions and truncate the Bessel series 
$\sigma_g(t)=\frac{t}{e^t-1}\sum_{n=1}^{M}{s_{n}(g)} \frac{J_n(2gt)}{2gt}$
with ${s_{n\ge M+1}}=0$.
The integral equation becomes a matrix equation
\begin{equation}
s(g)=\frac{1}{1+K(g)} \cdot h
\quad
\mbox{and}  
\quad
{\Gamma_{\rm cusp}(g)}=4 g^2 {s_{1}(g)},
\label{eq:mat}
\end{equation} 
with  $K(g)$ being a complicated $g-$dependent matrix
($M \times M$)
and $h=(1,0,0,0,0,\ldots)$ is a boundary condition vector. 
The numerical solution to Eq.~(\ref{eq:mat}) is presented in Fig.~\ref{fig:cusp}.

The first few coefficient of the strong coupling expansion was fitted 
in Ref.~\cite{Benna:2006nd}:
\begin{equation}
2 {\Gamma_{\rm cusp}(g)}=4.000000 {g} - 0.661907 - 0.0232 {g^{-1}}+\ldots,
\end{equation}
One can see that the first two coefficients agree with string theory, for instance $0.661907 \approx \ft{3 \ln 2}{\pi}$. 
From the numerical results two questions arise: if it is possible to calculated these coefficients from ${\cal N}=4$ SYM theory analytically and if the third coefficient is consistent with the string theory\footnote{Early calculations of A.~Tseytlin et al. disagree with this value.}.

\section{Ordinary strong coupling expansion }
One can try to perform the strong coupling expansion analytically
using the infinite Bessel series, \ie working in
infinite-dimension matrices
\cite{Alday:2007qfKostov:2007kxBeccaria:2007tk}:
\begin{equation}
\left[1+K(g)\right] \cdot s(g)=h,
\end{equation}
Expanding the matrix $K(g)$ and the solution $s(g)$ in powers of $1/g$ gives:
\begin{equation}
s_{n}(g)=g^{-1} \sum_{j=0} g^{-j} s_n^{(j)},
\quad
K(g)=g \sum_{j=0} g^{-j} K^{(j)},
\end{equation}
The leading order solution, $s^{(0)}=[K]^{-1} \cdot h+ [\mbox{zero modes}]$,
is defined up to zero modes.
This ambiguity is fixed in Ref.~\cite{Alday:2007qfKostov:2007kxBeccaria:2007tk} by the constraint from numerics
$s^{(0)}_{2k-1}=s^{(0)}_{2k}$,
$s^{(0)}_{2k-1}=s^{(0)}_{2k}=(-1)^{k+1}\ft{\Gamma(k+\frac1{2})}{\Gamma(k)
\Gamma(\frac1{2})}$,   
For the next-to-leading order constraints fixing zero modes 
are more complicated and cannot by easily extracted from numerics.

\begin{figure}
\begin{center}
{
\psfrag{n}[cc][cc]{$n$} \psfrag{sn}[cc][cc]{$ $}
\psfrag{25}[cc][cc]{\scriptsize $25$}
\psfrag{20}[cc][cc]{\scriptsize $20$}
\psfrag{15}[cc][cc]{\scriptsize $15$}
\psfrag{10}[cc][cc]{\scriptsize $10$}
\psfrag{5}[cc][cc]{\scriptsize $5$}
\psfrag{\2612}[cc][cc]{\scriptsize $-2$}
\psfrag{\2611}[cc][cc]{\scriptsize $-1$}
\psfrag{0}[cc][cc]{\scriptsize $0$}
\psfrag{1}[cc][cc]{\scriptsize $1$}
\psfrag{2}[cc][cc]{\scriptsize $2$}
\psfrag{exact}[ll][ll]{ $ s^{\rm (exact)}_n(g={\rm fixed})$}
\psfrag{PT}[ll][ll]{$ 
s^{(0)}_n+g^{-1}s^{(1)}_n$}
{\epsfysize4.5cm \epsfbox{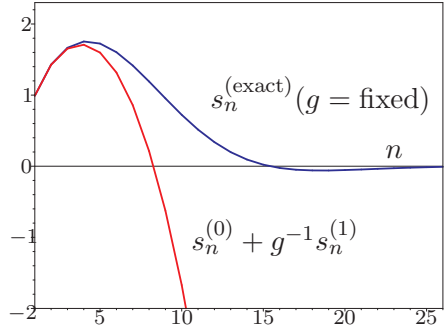}}
}
\end{center}
\caption{The $n-$dependence of  the $s_n(g)$ coefficients and its comparison
to a first few term of strong coupling expansion.
Both lines are plotted for large fixed $g$.} 
\label{fig:mismatch}
\end{figure}

\section{Our main idea }
Let us define the even and odd unknown:
\begin{equation}
s_{n}^\pm(g)=\ft{1}{2}(s_{2 n -1}(g)\pm s_{2n}(g)),
\end{equation}
From numerical calculations in Fig. \ref{fig:mismatch} one 
can see that the strong coupling expansion works well for  $n\sim1$
while it  fails for large $n$.

The approach proposed in Ref.~\cite{Basso:2007wd} reads as follows:
\begin{enumerate}
\item construct the solution for $s^{\pm}_n(g)$ in the region $n \sim 1$  and parameterize the contribution of (zero modes) by  yet unknown coefficients $c^{\pm}_{p}(g)$,
\item  construct the asymptotic solution for $s^{\pm}_n(g)$ in the region $n \gg 1$,
\item  sew two asymptotic expressions for $s^{\pm}_n(g)$ in the intermediate region $n \sim g^{1/2} $
 and determine the infinite  set of zero mode coefficients $c^{\pm}_{p}(g)$.
\end{enumerate}
One can do it without knowing the exact solution
performing the scaling limit 
$n,g \to \infty$  where $x=(n-\frac1{4})^2=$fixed.

\section{Fixing zero mode coefficients}

Changing variables properly \cite{Basso:2007wd} 
one can find a solution of the BES equation:
\begin{equation}
{\Gamma_{\rm cusp}(g)} = {2g}  + \sum_{p=1}^\infty \frac1{g^{p-1}}
\left[\frac{2 {c_p^-(g)}}{\sqrt{\pi}} {\Gamma(2p-\ft32)}
+ \frac{2 {c_p^+(g)}}{\sqrt{\pi}}
{\Gamma(2p-\ft12)} \right], 
\label{eq:cuspc}
\end{equation}
as a function of yet unknown 
${c_p^\pm(g)}= \sum_{r\ge 0} g^{-r} {c_{p,r}^\pm}$.
The expansion coefficients $s^\pm_m(g)$
should have correct the
scaling behavior in the  scaling limit
\begin{equation}
m,g \to \infty
\quad 
\mbox{for}
\quad
x=(m-\frac{1}{2})^2/g=\mbox{fixed},
\end{equation}
so that
\begin{equation}
s^\pm_m(g)=
\frac{(gx)^{-1/4}}{g \sqrt{\pi}}\left[\gamma_{\pm}^{(0)}(x)+
+\frac{\gamma_{\pm}^{(1)}(x)}{gx}+{\cal O}(1/g^2)
\right],
\end{equation}
where the expansion of
 $\gamma_{\pm}^{(r)}(x)$ 
runs in integer positive powers of $x$ and
$\gamma_{\pm}^{(r)}(x)$ 
should have a faster-than-power-law decrease at large $x$, or equivalently,
its Laplace transform should be an analytical function.

\section{ Quantization conditions }

Let us consider the leading order of $1/g$.
Performing the Laplace transform  from $\gamma_{\pm}^{(0)}(x)$ 
to $\tilde \gamma_{\pm}^{(0)}(s)$ 
one can obtain the analyticity condition 
\begin{equation}
{
{\sum_{p\ge 0} s^{p}\,{ c_{p,0}^+} \Gamma(p-\ft14)}
}
= 2[\Gamma(\ft34)]^2 \frac{\Gamma \left( 1-
{\frac {s}{2\pi}} \right)}{\Gamma  \left( \ft34- {\frac {s}{2\pi}} \right) }, 
\qquad
{c_{0,0}^+}=-\ft12,
\label{eq:q1}
\end{equation}
corresponding to cancellation roots and poles of
$\tilde \gamma_{+}^{(0)}(s)$.
A similar condition for $\tilde \gamma_{-}^{(0)}(s)$ reads as follows
\begin{equation}
{
{\sum_{p\ge 0} s^{p}\,
\left[{c_{p,0}^-}
\Gamma(p-\ft34) +
2{c_{p,0}^+}\lr{p-\ft14}{\Gamma(p+\ft14)}\right]}
}
 = {\frac{[\Gamma(\ft14)]^2}{4}}\frac{\Gamma \left( 1- {\frac {s}{2\pi}} \right)}{\Gamma \left( \ft14-{\frac {s}{2\pi}} \right) }
\label{eq:q2}
\end{equation}
 where ${c_{0,0}^-}=0$.
The expansion of Eqs. (\ref{eq:q1}) and (\ref{eq:q2}) in $s$ gives
\begin{equation}
{c_{1,0}^+} = - {\frac {3\ln 2 }{\pi }}+\frac12+ O(1/g)\,,
\quad {c_{1,0}^-} = {\frac {3\ln 2}{4\pi }}-\frac14+ O(1/g).
\label{eq:cc}
\end{equation}
Substituting (\ref{eq:cc}) to (\ref{eq:cuspc}) one gets
\begin{equation}
{\Gamma_{\rm cusp}(g)} = 2 {g} - \frac{3\ln 2}{2\pi} + O({g^{-1}}).
\end{equation}
Using this method one can continue the calculations 
for higher orders of $1/g$ series.
Finally, one gets
\begin{eqnarray}
{\Gamma_{\rm cusp}\left({g}{+c_1}\right)}& =&
2{g}\bigg[1- c_2 {g^{-2}}-
c_3 {g^{-3}}-\lr{ c_4+2\,c_2^{2}} {g^{-4}} 
\nonumber  \\
&&
- \lr{c_5 +23\,c_2c_3} {g^{-5}} - \lr{c_{{6}}+{\ft {166}{7}}
\,c_{{2}}c_{{4}} +54\,c_3^{2}+ 25\,c_2^{3}} {{g}^{-6}}
\nonumber  \\
&&
- \lr{ c_{{7}}+{\ft {1721}{29}}\,c_{{2}}c_{ {5}}+{\ft
{1431}{7}}\,c_{{3}}c_{{4}}+457\, c_{{2}}^{2}c_ {{3}}} {{g}^{-7}} 
\nonumber  \\
&&
  - \left(c_{{8}}+{\ft {6352}{107}}\,c_{{2}}c_{{6}}+{ \ft
{12606}{29}}\,c_{{3}}c_{{5}}+{\ft {7916}{49}} \, c_{{4}}^{2}+{\ft {6864}{7}}\,
c_{{2}}^{2}c_{ {4}} \right. 
\nonumber  \\
&&
 \left. +4563\,c_{{2}}c_{{3}}^{2} +374\, c_{{2}}^{4}\right)
{{g}^{-8}} 
\nonumber  \\
&&
-\left(
c_9
+\ft{30943}{277} c_7 c_2
+\ft{72089}{107} c_6 c_3
+\ft{216437}{203} c_5   c_4
\right. 
\nonumber  \\
&&
 \left.
+\ft{71712}{29} c_5 c_2^2
+17016 c_4 c_3 c_2
+13131 c_3^3
+16904 c_3 c_2^3
\right)
{{g}^{-9}} 
\nonumber  \\
&&
-\left(
c_{10}
+\ft{464314 }{4183} c_8 c_2
+\ft{308416}{277} c_7 c_3
+\ft{154466}{107} c_6 c_4
\right. 
\nonumber  \\
&&
 \left.+\ft{455267}{107} c_6 c_2^2
+\ft{1296500 }{841} c_5^2
+\ft{1818888}{29} c_5 c_3 c_2
+\ft{1137597}{49} c_4^2 c_2
\right. 
\nonumber  \\
&&
 \left. 
+\ft{756936}{7} c_4 c_3^2
+\ft{254648}{7} c_4 c_2^3
+253728 c_3^2 c_2^2
+10666 c_2^5
\right){{g}^{-10}} 
\nonumber  \\
&&
+O \left( {{g}^{-11}} \right)  \bigg],
\label{eq:cusp}
\end{eqnarray}
where the expansion coefficients are given by
\begin{equation}
  c_{{1}}=\ft{3\ln 2}{4\pi},\quad  c_{{2}}=\ft{1}{16\pi^2}\textrm{K}, \quad
c_{{3}}= \ft{{27}}{2^{11}\pi^3}\zeta(3),\quad  c_{{4}}=\ft{{21}}{2^{10}\pi^4}\beta(4),
\nonumber 
\end{equation}
\begin{equation}
c_{{5}}=\ft{{43065}}{2^{21}\pi^5}\zeta(5) ,\; c_{{6}}=
\ft{{1605}}{2^{15}\pi^6}\beta(6), \;
c_{{7}}=  \ft{{101303055}}{2^{30}\pi^7}\zeta(7), \;
\nonumber 
\end{equation}
\begin{equation}
 c_{{8}}=
\ft{{1317645}}{2^{22}\pi^8}\beta(8),\;
c_{{9}}= \ft{{1991809466325}}{2^{41}\pi^{9}}\zeta(9), \; 
c_{{10}}= \ft{{524012895}}{2^{27}\pi^{10}}\beta(10),
\end{equation}
with the Riemann zeta function, $\zeta(x)=\sum_{n\ge 1}{n^{-x}}$,
the Dirichlet beta function, $\beta(x)=\sum_{n\ge 0}{(-1)^n}{ (2n+1)^{-x}}$
and $\textrm{K}=\beta(2)$ the Catalan's constant. 
To simplify the formula the cusp in Eq. (\ref{eq:cusp}) is shifted by $c_1$.
One has to notice that 
all expansion coefficients, except the first one, are negative and they
decrease with the series order.


\section{ Asymptotic expansion ${\cal O}(g^{-40})$}

Using the above method one can evaluate numerically  the strong coupling expansion coefficients to rather high order and find that
the asymptotic expansion is not Borel summable 
\begin{equation}
 \Gamma_{\rm cusp}(g)\! \sim \! -g \sum_k \frac{\Gamma(k-\ft12)}{(2\pi g)^k} = g
\int_0^\infty \frac{du\, u^{-1/2}\e^{-u}}{u-2\pi g},
\end{equation}
where the Borel transform has a pole at $u=2\pi g$.

Ambiguity due to different prescriptions 
to integrate over the pole is   for large $g$
$\delta \Gamma_{\rm cusp }(g) \sim  g^{1/2} \exp (-2 \pi g)$.
Similar corrections  appear in  the solution
of the  FRS equation 
\cite{Freyhult:2007pzBasso:2008tx}
\begin{equation}
\gamma_{S}^{(L)}(g) =2 ({\Gamma_{\rm cusp}(g)}+ \epsilon(g,L) \ln S+ \ldots,
\end{equation}
This result agrees with the $O(6)$ sigma model from the string theory side
\cite{Alday:2007mf}.

\section{Conclusions}
The above calculations of the cusp anomalous dimension
of ${\cal N}=4$ SYM theory
shows that the integrability provides us strong methods 
for solving complicated problems.
Both, the weak and strong coupling expansion of the cusp 
can be found to an arbitrary order\footnote{The calculations are 
limit only by ability of programs for symbolic calculations}. 
The results agree with result 
from the string theory\footnote{
After our work A.~Tseytlin et. al. corrected their two loop
string results, which  now agrees with out results}
confirming validity
of the AdS/CFT correspondence \cite{Roiban:2007jf}.
Moreover,
it seems than on MSYM theory side  it is easier also to calculate
strong coupling expansion of the cusp anomalous dimension. 
Therefore,
confirmation of the above results from the string theory 
will be a challenging task.

\vspace{1cm}
I would like to warmly thank to G.~P.~Korchemsky and B.~Basso for the collaboration.
This work was supported by the grant of 
SFB 676, Particles, Strings and the Early Universe:
the Structure of Matter and Space-Time
and the grant of
the Foundation for Polish
Science.


\end{document}